\documentclass[amsmath,pre,twocolumn,superscriptaddress]{revtex4}  
\usepackage{wasysym}
\usepackage{graphicx}
\usepackage{xcolor}
\usepackage{amssymb}

\def\D{\mathrm{d}}

\begin{document} 
\title{Dynamically accelerated cover times} 

\author{Gcina Maziya}
\affiliation{Department of Mathematics, Imperial College London, 180 Queen's Gate, London SW7 2AZ, UK}
\affiliation{London Mathematical Laboratory, 8 Margravine Gardens, London W6 8RH, UK}
\affiliation{Centre for Complexity Science, Imperial College London, London SW7 2AZ, UK}
\author{Luca Cocconi}
\affiliation{Department of Mathematics, Imperial College London, 180 Queen's Gate, London SW7 2AZ, UK}
\affiliation{The Francis Crick Institute, 1 Midland Rd, London NW1 1AT, UK}
\affiliation{Centre for Complexity Science, Imperial College London, London SW7 2AZ, UK}
\author{Gunnar Pruessner}
\affiliation{Department of Mathematics, Imperial College London, 180 Queen's Gate, London SW7 2AZ, UK}
\affiliation{Centre for Complexity Science, Imperial College London, London SW7 2AZ, UK}
\author{Nicholas R. Moloney}
\affiliation{Department of Physics, Imperial College London, Prince Consort Road, London SW7 2AZ, UK}
\affiliation{Centre for Complexity Science, Imperial College London, London SW7 2AZ, UK}
\affiliation{Department of Mathematics and Statistics, University of Reading, Reading RG6 6AX, UK}

\begin{abstract}
Among observables characterizing the random exploration of a graph or
lattice, the cover time, namely the time to visit every site,
continues to attract widespread interest. Much insight about cover
times is gained by mapping to the (spaceless) coupon collector
problem, which amounts to ignoring spatiotemporal correlations, and
an early conjecture that the limiting cover time distribution of
regular random walks on large lattices converges to the Gumbel
distribution in $d \geq 3$ was recently proved
rigorously. Furthermore, a number of mathematical and numerical
studies point to the robustness of the Gumbel universality to
modifications of the \textit{spatial} features of the random search
processes (e.g., introducing persistence and/or intermittence, or
changing the graph topology). Here we investigate the robustness of
the Gumbel universality to dynamical modification of the
\textit{temporal} features of the search, specifically by allowing the
random walker to ``accelerate" or ``decelerate" upon visiting a
previously unexplored site. We generalize the mapping mentioned above
by relating the statistics of cover times to the roughness of
$1/f^\alpha$ Gaussian signals, leading to the conjecture that the
Gumbel distribution is but one of a family of cover time
distributions, ranging from Gaussian for highly accelerated cover, to
exponential for highly decelerated cover. While our conjecture is
confirmed by systematic Monte Carlo simulations in dimensions $d > 3$,
our results for acceleration in $d=3$ challenge the current
understanding of the role of correlations in the cover time problem.

\end{abstract} 

\date{\today}

\maketitle

\section{Introduction}

How long does it take to collect $N$ distinct objects that are sampled
uniformly with replacement? This is the so-called coupon collector
problem~\cite{Holst1986}. Depending on the context, the objects may
represent stickers in a football album, vertices on a fully connected
graph, or people in an epidemic. Close analogies to the coupon
collector can be found in a toy model for the buildup of strain in a
seismic fault~\cite{Gonzalez2005}, the random deposition of $k$-mers
on a substrate~\cite{Turban2019}, the infection of nodes on a
network~\cite{Ottino2017}, or the parasitization of
hosts~\cite{Zoroa2017}. More generally, the coupon collector belongs
to the family of urn problems~\cite{JohnsonKotz1977,Holst2001}. An
early result, proved by Erd\H{o}s and R\'enyi~\cite{Erdos1961}, is
that the coupon collection time follows a Gumbel distribution.

Often, the $N$ objects to be collected are not sampled uniformly at
any given time. For example, a random walker exploring a lattice can
only ``collect'' nearest-neighbor sites. In this context, the total
time to visit every site on a graph or lattice is known as the cover
time. Cover times have been intensely studied since the
1980s~\cite{Aldous1983,Aldous1989,Wilf1989}. For example, an early
conjecture~\cite{Aldous_book1989} that the cover time for a $d\ge3$
torus is also Gumbel distributed was recently proved
rigorously~\cite{Belius2013}. The manner in which a random walker
covers a lattice~\cite{Brummelhuis1991,Brummelhuis1992,Freund1993} is
encoded in the trace of the walk, i.e., the walk's history, and this
nontrivial random object has received much attention in the
mathematics literature~\cite{Sznitman2012,Drewitz2014}. Qualitatively,
an important distinction is between walks that are transient ($d > 2$)
versus recurrent ($d \le 2$), even if the walk is restricted to a
finite torus, in which case every site will eventually be visited.

In this paper, we are interested in modifying the cover process in
time. Thus, we study the consequences of accelerating or decelerating
the random walker upon visiting a new site. In this way, we show that
the Gumbel distribution is but one of a family of cover time
distributions, ranging from Gaussian for highly accelerated cover, to
exponential for highly decelerated cover. Coincidentally, this family
of distributions describes the roughness of $1/f^{\alpha}$ Gaussian
signals~\cite{Antal2002}.

Our motivation for dynamically modifying the cover process is to
further investigate some of the assumptions underlying the mapping of
the cover time problem in $d \geq 3$ to the coupon collector problem,
specifically those relating to the irrelevance of spatiotemporal
correlations. The specific procedure we implement is also inspired by
transport behavior in, e.g., cellular environments, in which a molecule
may aggregate or fragment in the course of its diffusion, thereby
altering its diffusion constant in
time~\cite{Coquel2013,Soria2019}. Alternatively, in the context of
search problems~\cite{Benichou2011}, the random walker could be
``rewarded'' or ``penalized'' upon acquiring new targets, thereby
enhancing or inhibiting future search.

The structure of the paper is as follows: In Sec.~\ref{S:CCP} we
review basic results of the coupon collector problem. In
Sec.~\ref{S:CCPmod} we describe how we accelerate or decelerate the
dynamics, and identify the distribution of collection times. In
Sec.~\ref{S:torus} we turn our attention to cover times on a torus,
and present numerical results for accelerated and decelerated random
walkers in Secs.~\ref{S:celeration} and \ref{S:breakdown}. We
summarize our findings in Sec.~\ref{S:conclusion}.

\section{Coupon collector problem} \label{S:CCP}

In this section we review the basic properties of the coupon collector
problem~\cite{Erdos1961}. The probability $p_i$ of collecting a new
coupon, given that $i$ have already been collected, is
\begin{equation}
    p_i = 1-i/N, \quad i= 0, 1, \ldots, N-1.
\end{equation}
Qualitatively, the first coupons are collected rapidly, while the last
coupons are collected very slowly. Let $n_i$ be the number of coupons
drawn between collecting the $i$th and $(i+1)$th distinct coupon. Then
the total number of draws $C_N$ to collect $N$ coupons is
\begin{equation}
    C_N = \sum_{i=0}^{N-1} n_i,
\end{equation}
where $n_i$ are independent but nonidentical geometric random
variables with mean $1/p_i$. Using angular brackets to denote
expectation, the mean of $C_N$ is therefore
\begin{align}
    \langle C_N \rangle &= \sum_{i=0}^{N-1} \langle n_i \rangle \\
    &= \sum_{i=0}^{N-1} \frac{1}{1-i/N} = N\sum_{k=1}^{N} \frac{1}{k},
    \label{E:discrete}
\end{align}
which behaves like $N\ln N$ for large $N$, i.e. collecting the full
set of $N$ coupons is slower than linear in $N$. Similarly, it can be
shown that the variance of $C_N$ is proportional to $N^2$. Erd\H{o}s
and R\'enyi derived the full distribution of $C_N$, showing it to be
a Gumbel distribution~\cite{Erdos1961}.

Before giving a heuristic derivation of this distribution, it is
convenient to embed the coupon collector in continuous time, such that
coupons arrive at unit rate in the manner of a Poisson point
process~\cite{Aldous_book1989}. Thus, rather than the discrete unit
steps representing the number of coupon draws, consider instead the
amount of continuous time elapsed since collection began. In this
perspective, the collection time $H_j$ for any particular coupon $j$
is an exponential random variable with mean $N$,
\begin{equation}
    \mathbb{P}(H_j \le t) = 1-\exp(-t/N).
\end{equation}
The total collection time is the maximum of all the individual coupon
collection times. Since these times are identical and independent,
\begin{align}
    \mathbb{P}(C_N \le t) &= \mathbb{P}(\text{max} \{H_1, H_2, \ldots, H_N\} \le t) \label{E:hitting} \\
    &= \mathbb{P}(H_1\le t, H_2 \le t, \ldots, H_N \le t) \\
    &= \mathbb{P}(H_1\le t)\mathbb{P}(H_2 \le t)\cdots\mathbb{P}(H_N \le t) \\
    &= [1-\exp(-t/N)]^N \\
    &\to \exp[-N\exp(-t/N)], \quad\text{as $N \to \infty$.}
\end{align}
After centering and rescaling, 
\begin{equation}
    \mathbb{P}\left(\frac{C_N - N\ln N}{N} \le t\right) = \exp(-\exp(-t)),
\end{equation}
which is recognized as the Gumbel distribution from extreme value
statistics~\cite{Leadbetter1983}.

\section{Accelerated and decelerated coupon collector} \label{S:CCPmod}

The waiting time $T_i$ between collecting the $i$th and $(i+1)$th
distinct coupon is a sum over a random number $n_i$ of unit
exponential random variables. Since $n_i$ is a geometric random
variable, $T_i$ is, in fact, also exponentially distributed with mean
$1/p_i$~\cite{Cox1980}. Thus, the total collection time can be written
as
\begin{equation}
    C_N = \sum_{i=0}^{N-1} T_i = N \sum_{k=1}^{N} \frac{\varepsilon_k}{k},
    \label{E:sum}
\end{equation}
where $\varepsilon_k$ are independent and identically (i.i.d.)
distributed exponential random variables with unit mean.

\begin{figure}
    \centering
    \includegraphics[width=\columnwidth]{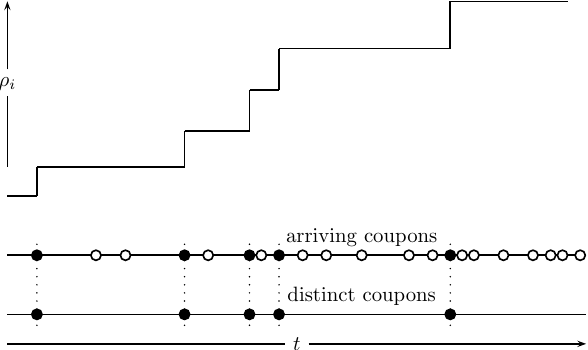}
    \caption{Schematic illustration of an acceleration protocol. The
      intensity of coupon arrivals (middle) is increased as distinct
      coupons are acquired (solid circles). The piecewise constant and
      increasing intensity profile (top) gives rise to a point process
      of distinct coupon arrivals (bottom) whose intensity can be
      adjusted.}
    \label{F:acceleration}
\end{figure}

We now manipulate the arrival rate of random coupons which, in turn,
alters the rate at which distinct coupons are collected. For example,
if coupons arrive at rate $\rho_i = 1/p_i = 1/(1-i/N)$ all the while
$i$ coupons have been collected, then the waiting time between
distinct coupons has unit mean. Thus, by accelerating the arrival of
coupons to compensate for the decreasing likelihood of obtaining a
distinct coupon, distinct coupons are collected at unit rate. This
acceleration protocol is depicted schematically in
Fig~\ref{F:acceleration}: the piecewise constant rates $\rho_i$
increase each time a distinct coupon is collected.

In order to accommodate a variety of acceleration-deceleration
protocols, we generalize the rates $\rho_i$ according to
\begin{equation}
    \rho_i(\alpha) = p_i^{\alpha-1}, \quad \alpha \ge 0.
    \label{E:rescale}
\end{equation}
This leads to the collection time
\begin{equation}
    C_N(\alpha) = N^{\alpha}\sum_{k=1}^N \frac{\varepsilon_k}{k^{\alpha}},
\label{E:CNalpha}
\end{equation}
where the unaccelerated coupon collector is recovered for $\alpha =
1$, i.e. Eq.~\eqref{E:sum}, and the accelerated version just discussed
above corresponds to $\alpha = 0$. For large $N$, the mean of
$C_N(\alpha)$ scales as
\begin{equation}
    \langle C_N(\alpha) \rangle \sim \begin{cases} N, &\quad 0 \le \alpha<1 \\
    N \ln N, &\quad \alpha = 1\\
    N^{\alpha}, &\quad \alpha > 1,
    \end{cases}
\end{equation}
so that coupon collecting is accelerated for $0 \le \alpha<1$, and
decelerated for $\alpha >1$, as compared to the original unaccelerated
process with $\alpha = 1$.

Apart from the $N^{\alpha}$ prefactor, the exact same sum in
Eq.~\eqref{E:CNalpha} describes the roughness of periodic Gaussian
$1/f^{\alpha}$ signals~\cite{Antal2002}, as outlined in
Appendix A. In that context, $\alpha = 0,1,2,4$ correspond respectively
to white noise, $1/f$ noise~\cite{Weissman1988}, a steady-state
Edwards-Wilkinson interface~\cite{Edwards1982}, and a steady-state
curvature-driven interface~\cite{Mullins1957}.

When $\alpha = 0$, $C_N(0)$ in Eq.~\eqref{E:CNalpha} is a sum over
independent and identically distributed random exponential variables,
which, after rescaling, is described by the central limit theorem. As
shown in Appendix B, the Lindeberg condition extends the central
limit theorem to nonidentical random variables, such that the
rescaled distribution of $C_N(\alpha)$ remains Gaussian for all
$\alpha \le 1/2$. For $\alpha = 2$, the distribution is
Kolmogorov-Smirnov, i.e. the distribution of the test statistic in the
Kolmogorov-Smirnov goodness-of-fit test~\cite{Kolmogorov1933}. This
distribution reoccurs in many Brownian
problems~\cite{Foltin1994,Biane2001}, branching
processes~\cite{Font2016}, aggregation~\cite{Botet2005}, and
statistics~\cite{Watson1961}. For $\alpha = 4$, the distribution of
$C_N(4)$ has been calculated in~\cite{Plischke1994}. Finally, in the
limit $\alpha \to \infty$, $C_N(\infty)$ is exponentially distributed,
since only the first term in Eq.~\eqref{E:CNalpha} contributes. A full
discussion of the properties of $C_N(\alpha)$ can be found
in~\cite{Antal2002}. In summary, the Gumbel distribution is one of a
family of distributions of sums of weighted exponential random
variables.

\section{Cover times on a torus} \label{S:torus}

\begin{figure}[h!]
    \centering
    \includegraphics[width=\columnwidth]{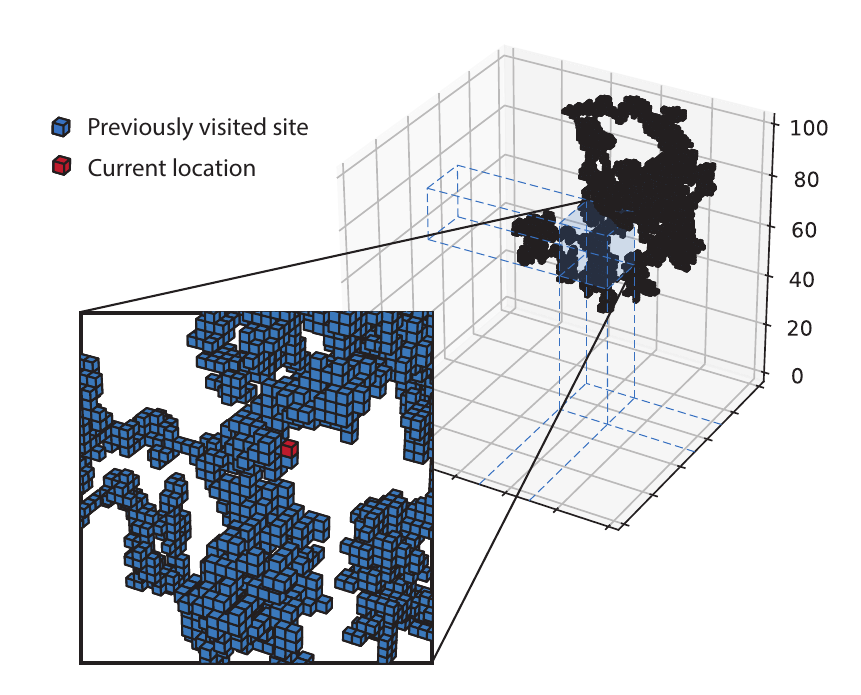}
    \caption{(Color online) Portion of the trace of a random walk in
      $d=3$, showing the sites visited as the walker covers the
      lattice.}
    \label{F:trace}
\end{figure}

If one identifies coupons with sites, then coupon collecting is
similar in spirit to covering a lattice or graph, that is, visiting
each and every site at least once. However, if the lattice exploration
is undertaken by a random walker, it is far from obvious that coupon
collecting describes the statistics of covering: at any given time
coupons are sampled uniformly, whereas a random walker samples
nearest-neighbor sites. This nonuniform sampling is illustrated in
Fig.~\ref{F:trace}, showing a portion of the trace of a random walk as
it covers a lattice in $d=3$.

On a fully connected graph all sites are nearest neighbors. Therefore,
covering a fully connected graph via a random walk is almost identical
to coupon collecting, with the irrelevant difference that the random
walker must necessarily leave the site most recently visited (assuming
self-loops are excluded). Meanwhile, for random graphs cover times
have been actively studied by
mathematicians~\cite{Aldous1989,AldousFill} and
physicists~\cite{Zlatanov2009,Maier2017}, among others. If the
probability distribution of the random walker location converges to
the uniform distribution sufficiently fast, the same $N\ln N$ scaling
as Eq.~\eqref{E:discrete} often describes the mean cover time. A
graph-dependent constant prefactor will reflect the fact that the
walker has to diffuse across the graph to cover it. This constant can
be expressed in terms of the mean time spent at the
origin~\cite{Aldous1983}.

For random walks on a torus (i.e., a regular lattice with periodic
boundary conditions), cover times depend on dimension. In $d=1$, the
cover time (equivalent to the first-passage time of the range process)
is not Gumbel distributed~\cite{Imhof1985}, while in $d\ge 3$ it
is~\cite{Belius2013}. The $d=2$ cover time, posed as the ``white
screen problem"~\cite{Wilf1989}, is not completely resolved to this
day. Dembo \textit{et al.} have established rigorously that the mean
cover time converges to $4L^2 (\ln L)^2/\pi$ as the side length $L$ of
the simple cubic lattice tends to infinity, although there are
practical difficulties in observing this behavior in
numerics~\cite{Grassberger2017}. Subleading order corrections to Dembo
\textit{et al.}'s result have been explored in the mathematics
literature~\cite{Belius2017}. In the physics literature, numerical
evidence suggests that $d=2$ cover times are approximately
Gumbel distributed~\cite{Chupeau2015}.

For this reason, in the following we restrict our attention to $d\ge
3$, where it is rigorously known that the cover time is Gumbel
distributed~\cite{Belius2013} (already anticipated heuristically
in~\cite{Aldous_book1989}). The technical proof of this result relies
on the transience of a random walker in $d\ge 3$, and the
approximately Poisson distribution of unvisited sites at the late
stage of the cover process~\cite{Belius2013}. Remarkably, the coupon
collector scaling carries over to the cover time, even though the
first-passage times $\{H_1,H_2,\ldots,H_N\}$ to each of the $N$ sites
are clearly not independent random variables, although they are
approximately exponential. The appropriately scaled cover time now
takes the form
\begin{equation}
    \frac{C_N - g(0) N\ln N}{g(0)N},
\end{equation}
which is identical to the coupon collector apart from a factor
$g(0)$. This factor is the Green function for the unrestricted random
walker evaluated at the origin, which is equivalent to the mean time
spent at the origin. For example, for the simple cubic lattice in
$d=3$~\cite{Glasser1977}
\begin{align}
    g(0) &= \frac{4\sqrt{6}}{\pi^2}\Gamma\left(\frac{1}{24}\right)
    \Gamma\left(\frac{5}{24}\right)\Gamma\left(\frac{7}{24}\right)
    \Gamma\left(\frac{11}{24}\right) \notag \\
    &= 1.516...
\end{align}
Thus, random walk covering is approximately $50\%$ slower on a
simple cubic lattice compared to a fully connected graph.

\section{Accelerated and decelerated cover \label{S:celeration}}

In the coupon collector, the waiting times between coupon arrivals are
exponential, and acceleration or deceleration is effected by changing
its rate. Analogously, the cover process is accelerated or decelerated
by changing the rate of the exponential waiting times between random
walk steps. Thus, if we employ the acceleration-deceleration protocol
as described in Eq.~\eqref{E:rescale}, we might conjecture that, for
$d\ge 3$, the cover time in Eq.~\eqref{E:CNalpha} is generalized to
\begin{equation}
    C_N(\alpha) = g(0) N^{\alpha} \sum_{k=1}^N \frac{\varepsilon_k}{k^\alpha},
\label{E:CNalpha_cover}
\end{equation}
where $\varepsilon_k$ are again i.i.d. exponential random variables,
and the effect of the underlying lattice is incorporated by the Green
function $g(0)$. This generalization assumes that the correlations
that were carefully accounted for in the standard cover problem
\cite{Belius2013} continue to play a minor role for $\alpha \neq
1$. In the case $\alpha = 1$, it is known that the first-passage times
$H_\mathbf{x}$ and $H_\mathbf{y}$ of sites $\mathbf{x}$ and
$\mathbf{y}$, respectively, are correlated such that
\begin{equation}
    \text{Cov}(\mathbf{1}(H_\mathbf{x}>t),\mathbf{1}(H_\mathbf{y}>t)) \sim |\mathbf{x}-\mathbf{y}|^{-(d-2)}, \quad d \ge 3,
\label{E:correlations}
\end{equation}
where $\mathbf{1}(H_\mathbf{x}>t)$ indicates that site $\mathbf{x}$
has been visited at a time greater than $t$.
Equation~\eqref{E:correlations} is an asymptote in large system size $N$
with $t$ proportional to that
size~\cite{Brummelhuis1992,Drewitz2014}. For $\alpha \neq 1$, on the
other hand, the nature of the correlations is unknown to us.

We numerically test the conjecture of Eq.\eqref{E:CNalpha_cover} in
the following by rescaling the observed probability density
$p(C_N(\alpha))$ by the mean
\begin{equation}
\phi_1(x) = \langle C_N(\alpha) \rangle \, p\left(x \langle C_N(\alpha) \rangle \right)
\end{equation}
or by the standard deviation after centering,
\begin{equation}
    \phi_2(z)=
    \sigma_{C_N(\alpha)} \, p\left(z \sigma_{C_N(\alpha)} + \langle C_N(\alpha) \rangle \right)\ .
\end{equation}

\subsection{Deceleration, $\alpha = 2$}

\begin{figure}[h!]
    \centering
    \includegraphics[width=\columnwidth]{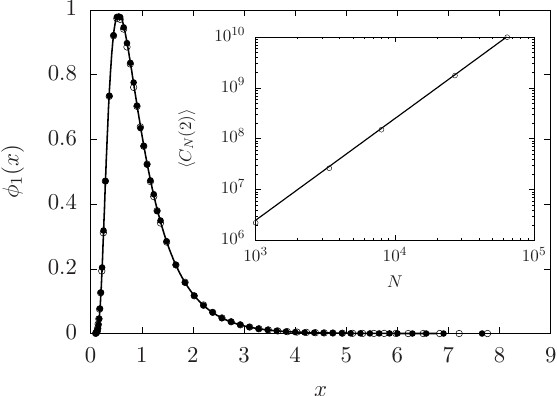}
    \caption{Rescaled cover time density, $\phi_1(x)$, for $\alpha=2$
      in $d=3$ with $N=30^3$ (open circles), and $d=4$ with $N=15^4$
      (solid circles), over an ensemble of $10^6$ independent
      realizations. The conjectured density (solid line) is given by
      Eq.~\eqref{E:a2theory}. Inset: Scaling of moments $\langle
      C_N(2)\rangle$ in $d=3$ (open circles). The conjectured behavior
      (solid line) is given by Eq.~\eqref{E:a2moments}. Standard
      errors are smaller than the symbols.}
    \label{F:a2lin}
\end{figure}

\begin{figure}[h!]
    \centering
    \includegraphics[width=\columnwidth]{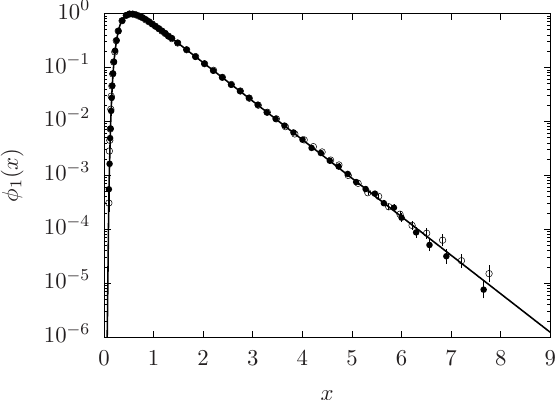}
    \caption{Same as Fig~\ref{F:a2lin} but with a logarithmic
      $y$axis. Error bars denote standard errors of histogram bins.}
    \label{F:a2loglin}
\end{figure}

For $\alpha = 2$, we conjecture that $C_N(2)$ is described by the
Kolmogorov-Smirnov distribution, with Laplace
transform~\cite{Foltin1994,Biane2001}
\begin{equation}
    \langle e^{-s C_N(2)} \rangle = 
    \frac{\sqrt{\pi^2 g(0)N^2 \, s}}{\sinh{\sqrt{\pi^2 g(0) N^2 \, s}}}
\label{E:a2rough}
\end{equation}
for large $N$, and first two moments
\begin{equation}
    \langle C_N(2) \rangle = \frac{\pi^2 g(0)}{6} \, N^2, \quad
    \langle C_N^2(2) \rangle = \frac{7 \pi^4 g^2(0)}{180} \, N^4.
    \label{E:a2moments}
\end{equation}
The Laplace transform in Eq.~\eqref{E:a2rough} can be inverted to
recover a series expansion for the probability density $p(C_N(2))$
which, after rescaling by the mean, reads~\cite{Foltin1994}
\begin{equation}
    \phi_1(x) = 
    \frac{\pi^2}{3} \sum_{k=1}^N (-1)^{k+1} k^2 \exp(-\pi^2 k^2 x/6).
    \label{E:a2theory}
\end{equation}
The sum converges fast, so that the cover time density of relatively
small systems is very close to the asymptotic density as $N\to
\infty$. Figures~\ref{F:a2lin} and~\ref{F:a2loglin} show excellent
agreement between empirical cover time densities and
Eq.~\eqref{E:a2theory} in $d=3,4$.

\subsection{Deceleration, $\alpha = 4$}

\begin{figure}[h!]
    \centering
    \includegraphics[width=\columnwidth]{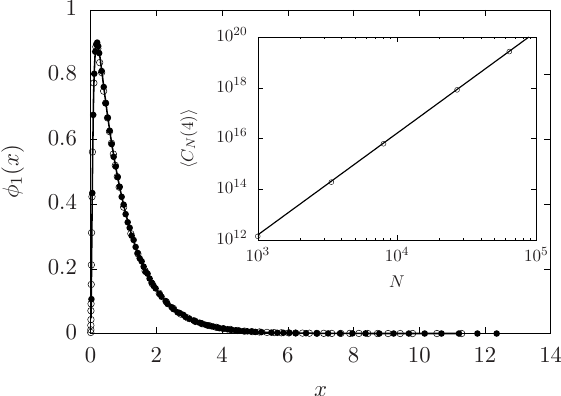}
    \caption{Rescaled cover time density, $\phi_1(x)$, for $\alpha=4$
      in $d=3$ with $N=30^3$ (open circles), and $d=4$ with $N=15^4$
      (solid circles), over an ensemble of $10^6$ independent
      realizations. The conjectured density (solid line) is given by
      Eq.~\eqref{E:a4theory}. Inset: Scaling of moments $\langle
      C_N(4)\rangle$ in $d=3$ (open circles). The conjectured behavior
      (solid line) is given by Eq.~\eqref{E:a4moments}. Standard
      errors are smaller than the symbols.}
    \label{F:a4lin}
\end{figure}

\begin{figure}[h!]
    \centering
    \includegraphics[width=\columnwidth]{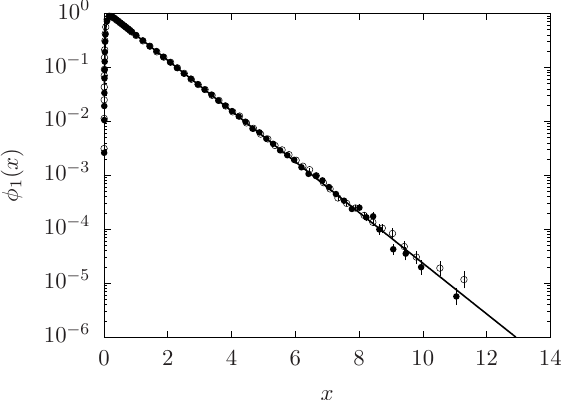}
    \caption{Same as Fig~\ref{F:a4lin} but with a logarithmic
      $y$-axis. Error bars denote standard errors of histogram bins.}
    \label{F:a4loglin}
\end{figure}

For $\alpha = 4$, we conjecture that $C_N(4)$ has the same
distribution as the roughness of a curvature-driven interface, with
Laplace transform~\cite{Plischke1994}
\begin{align}
    &\langle e^{-s C_N(4)} \rangle = \sqrt{4\pi^4g(0)N^4 s} \notag \\
    &\times \, \frac{1}{\sinh[(4\pi^4g(0)N^4s)^{1/4}]
    -\cos[(4\pi^4g(0)N^4 s)^{1/4}]}
\label{E:a4rough}
\end{align}
and first two moments
\begin{equation}
    \langle C_N(4) \rangle = \frac{\pi^4g(0)}{90} \, N^4, \quad
    \langle C_N^2(4) \rangle = \frac{13\pi^8g^2(0)}{56700} \, N^8.
    \label{E:a4moments}
\end{equation}
The Laplace transform in Eq.~\eqref{E:a4rough} can be inverted to
recover a series expansion for the probability density $p(C_N(4))$
which, after rescaling by the mean, reads~\cite{Plischke1994}
\begin{equation}
    \phi_1(x) = 
    \frac{2\pi^5}{45} \sum_{k=1}^N \frac{(-1)^{k+1} k^5}{\sinh(\pi k)} \exp(-\pi^4 k^4 x/90).
    \label{E:a4theory}
\end{equation}
Figures~\ref{F:a4lin} and~\ref{F:a4loglin} show excellent agreement
between empirical cover time densities and Eq.~\eqref{E:a4theory} in
$d=3,4$.

\subsection{Acceleration, $0 \le \alpha \le 1/2$, $d \ge 4$}

\begin{figure}[h!]
    \centering
    \includegraphics[width=\columnwidth]{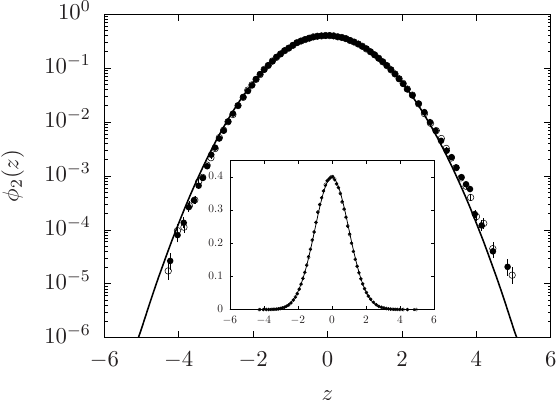}
    \caption{Rescaled cover time density, $\phi_2(z)$, for $\alpha=0$
      (open circles) and $\alpha=1/4$ (solid circles) in $d=4$ with
      $N=15^4$, over an ensemble of $10^6$ independent realizations,
      compared with the Gaussian conjecture (solid line). Error bars
      denote standard errors of histogram bins. Inset: $\phi_2(z)$ on
      linear axes.}
    \label{F:a0d4}
\end{figure}

\begin{figure}[h!]
    \centering
    \includegraphics[width=\columnwidth]{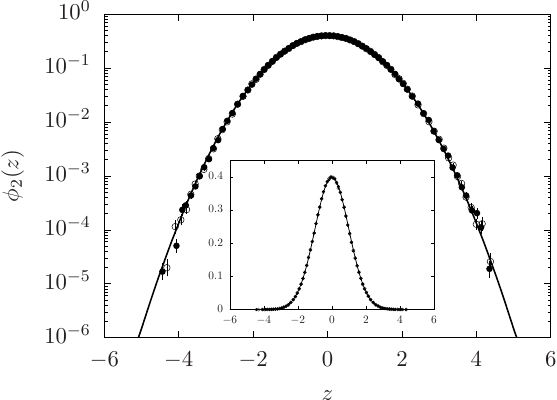}
    \caption{Rescaled cover time density, $\phi_2(z)$, for $\alpha=0$
      (open circles) and $\alpha=1/4$ (solid circles) in $d=5$ with
      $N=10^5$, over an ensemble of $10^6$ independent realizations,
      compared with the Gaussian conjecture (solid line). Error bars
      denote standard errors of histogram bins. Inset: $\phi_2(z)$ on
      linear axes.}
    \label{F:a0d5}
\end{figure}

For $0 \le \alpha \le 1/2$, Eq.~\eqref{E:CNalpha_cover} falls under
the scope of the central limit. Therefore, the conjectured statistics
of $C_N(\alpha)$ normalized to zero mean and unit standard deviation
are described by a Gaussian distribution
\begin{equation}
    \phi_2(z) = \frac{1}{\sqrt{2\pi}} e^{-z^2/2}.
\end{equation}
In the presence of correlations, the central limit theorem need no
longer apply. Indeed, we find that our conjecture breaks down for
accelerated cover in $d=3$, and we discuss that case separately in
Sec.~\ref{S:breakdown}. For $d \ge 4$, however, our conjecture
continues to agree well with numerics. Figure~\ref{F:a0d4} shows
empirical cover time densities for $\alpha = 0, 1/4$ and $d=4$. The
small asymmetric discrepancies from Gaussian behavior in the tails in
$d=4$ (Fig.~\ref{F:a0d4}) disappear altogether in $d=5$, as seen in
Fig~\ref{F:a0d5}. This is in keeping with the general notion that
correlations weaken with increasing dimension --- also suggested by
Eq.~\eqref{E:correlations}.

\subsection{Acceleration, $\alpha = 3/4$}

\begin{figure}[h!]
    \centering
    \includegraphics[width=\columnwidth]{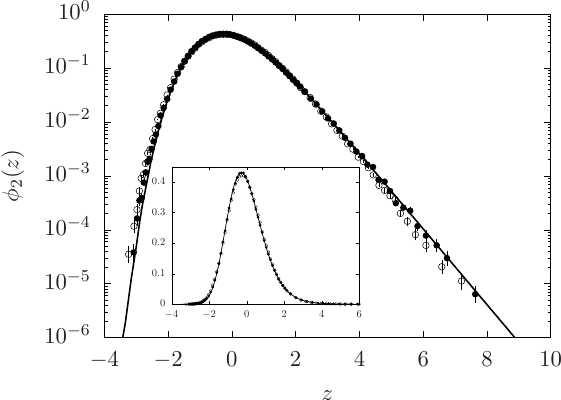}
    \caption{Rescaled cover time density, $\phi_2(z)$, for
      $\alpha=3/4$ in $d=3$ with $N = 30^3$ (open circles), and $d=4$
      with $N=15^4$ (solid circles), over an ensemble of $10^6$
      independent realizations. The conjectured density (solid line)
      is given by Eq.~\eqref{E:atheory}. Error bars denote standard
      errors of histogram bins. Inset: $\phi_2(z)$ on linear axes.}
    \label{F:a0.75}
\end{figure}

As explained in~\cite{Antal2002}, for $1/2 < \alpha < 1$ the rescaled
cover time densities $\phi_2(z)$ can be expanded as
\begin{equation}
    \phi_2(z) = \sqrt{\zeta(2\alpha)} \sum_{k=1}^{\infty}
    k^{\alpha} Y(\alpha,k) \exp(-k^{\alpha}\sqrt{\zeta(2\alpha)} \, z - 1), 
\label{E:atheory}
\end{equation}
where $\zeta$ is the Riemann zeta function, and
\begin{equation}
    Y(\alpha,k) = \prod_{n=1,\neq k}^{\infty} 
    \frac{e^{(-k/n)^{\alpha}}}{1-(k/n)^{\alpha}}.
\end{equation}
Equation~\eqref{E:atheory} defines an $\alpha$-dependent family of
distributions with exponential right tails. For the representative
case of $\alpha = 3/4$, Fig~\ref{F:a0.75} shows good agreement with
our conjecture, with small discrepancies in $d=3$ disappearing
altogether in $d=4$.

\section{Acceleration, $\alpha = 0$, $d=3$}\label{S:breakdown}

\begin{figure}[h!]
    \centering
    \includegraphics[width=\columnwidth]{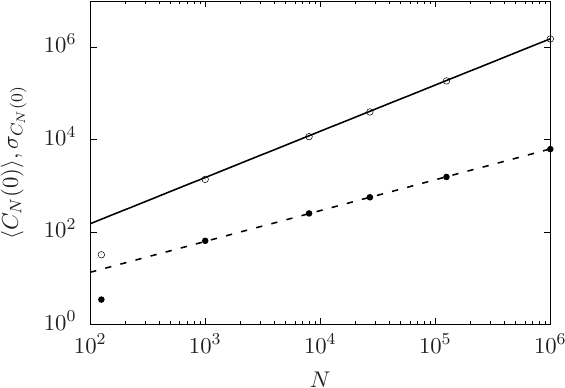}
    \caption{Scaling of the mean $\langle C_N(0)\rangle$ (open
      circles) and standard deviation $\sigma_{C_N(0)}$ (solid
      circles) for $\alpha = 0$ in $d=3$. The mean follows the
      conjectured behavior $g(0)N$ (solid line), but the standard
      deviation appears to scale as $N^{2/3}$ [dashed line,
      Eq.~\eqref{E:a0d3sd}], rather than $N^{1/2}$. Standard errors
      are smaller than the symbols.}
    \label{F:TWmusigma}
\end{figure}

In all cases considered so far, the conjecture that the cover time
$C_N(\alpha)$ is described statistically by
Eq.~\eqref{E:CNalpha_cover} is successfully verified
empirically. However, the conjecture fails in the case $\alpha = 0$ in
$d=3$. According to Eq.~\eqref{E:CNalpha_cover}, the cover time is
predicted to be statistically equivalent to a sum of independent and
identical exponential random variables, therefore falling under the
scope of the central limit theorem. The only feature correctly
predicted by Eq.~\eqref{E:CNalpha_cover} is that the mean cover time
$\langle C_N(0) \rangle$ still behaves as $g(0) N$, as shown in
Fig~\ref{F:TWmusigma}. However, the standard deviation
$\sigma_{C_N(0)}$ does not scale as $N^{1/2}$. Instead, for system
sizes $N \geq 10^3$ it is well approximated by
\begin{equation}
    \sigma_{C_N(0)} = A g(0) N^{\gamma},
\label{E:a0d3sd}
\end{equation}
where we note that the fitted values of the amplitude $A = 0.44(2)$
and exponent $\gamma = 0.6608(12)$ are close to $\sqrt{2}-1 =
0.414\ldots$ and $2/3$, respectively.

\begin{figure}[h!]
    \centering
    \includegraphics[width=\columnwidth]{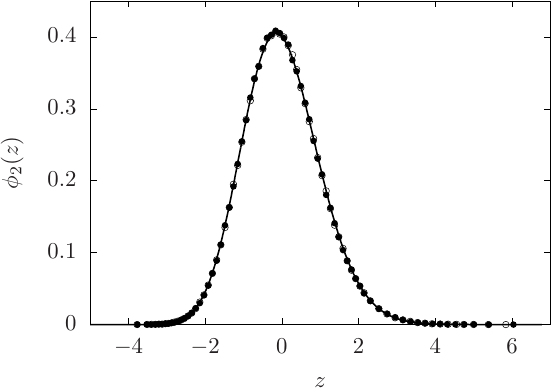}
    \caption{Rescaled cover time density, $\phi_2(z)$, for $\alpha=0$
      in $d=3$ with $N = 100^3$ (open circles) and $N=50^3$ (solid
      circles), over an ensemble of $10^6$ independent realizations. A
      Tracy-Widom density from the Gaussian orthogonal ensemble is
      plotted for comparison (solid line).}
    \label{F:TWlin}
\end{figure}

\begin{figure}[h!]
    \centering
    \includegraphics[width=\columnwidth]{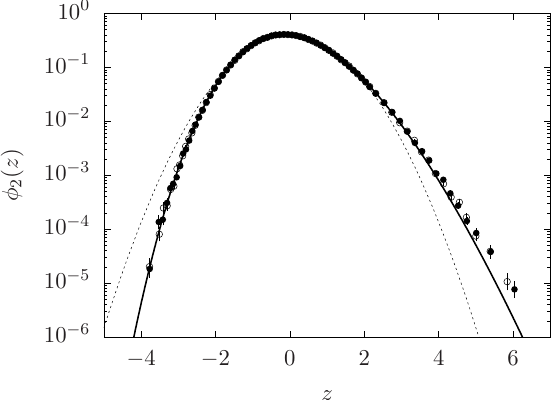}
    \caption{Same as Fig~\ref{F:TWlin} but with a logarithmic
      $y$ axis. Error bars denote standard errors of histogram
      bins. For comparison, a Gaussian density is also plotted (dotted
      line).}
    \label{F:TWloglin}
\end{figure}

\begin{figure}[h!]
    \centering
    \includegraphics[width=\columnwidth]{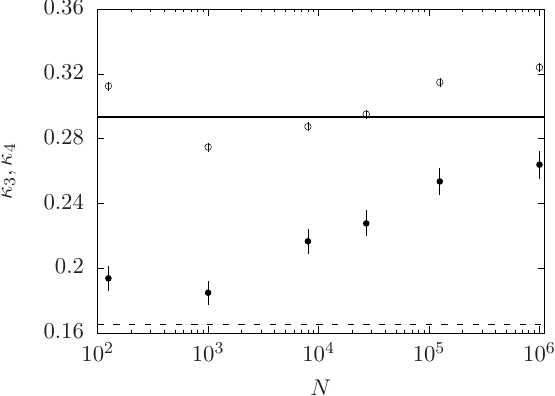}
    \caption{Skewness $\kappa_3$ (open circles) and kurtosis
      $\kappa_4$ (solid circles) for $\alpha = 0$ in $d=3$. Error bars
      denote jackknife standard errors. The Tracy-Widom skewness
      (solid line) and kurtosis (dashed line) are plotted for
      comparison.}
    \label{F:TWskewkurt}
\end{figure}

The rescaled cover time density $\phi_2(z)$ is also not Gaussian, as
shown in Figs~\ref{F:TWlin} and \ref{F:TWloglin}. We are not able to
identify the empirical density, although a Tracy-Widom density for the
largest eigenvalue from the Gaussian orthogonal ensemble of random
matrices gives a reasonable approximation. Given the discrepancies in
the right tail of the density, and the behavior of the skewness and
kurtosis as shown in Fig.~\ref{F:TWskewkurt}, we cannot claim
conclusive support for the Tracy-Widom density and offer this curious
near coincidence as an open problem.

While we cannot identify the empirical density of cover times for
$\alpha = 0$ and $d=3$, we can nevertheless investigate the breakdown
of our conjecture, Eq~\eqref{E:CNalpha_cover}, which naively expresses
the cover time as a sum over exponential waiting times
$\varepsilon_k$. Since we do not recover the anticipated Gaussian
distribution, we are led to conclude that the random variables
$\varepsilon_k$ are either sufficiently nonidentical, or
nonindependent (or both).

To isolate this question, we perform a shuffling operation across
(independent) members of the ensemble from which we collect statistics
of cover times. Specifically, we choose one member of the ensemble at
random, i.e., one realization of the cover process, and sum the first
of its $b$ waiting times $\{\varepsilon_k^{(1)}\}_{k=1}^b$. Then we
pick another realization at random, and sum the next $b$ waiting times
from that process $\{\varepsilon_k^{(2)}\}_{k=b+1}^{2b}$, and so
on. We continue this operation $N/b$ times, so that we accumulate the
shuffled cover time process
\begin{equation}
    C_{N}^{\text{shuff.}}(0) = \sum_{k=1}^b \varepsilon_k^{(1)} 
    + \sum_{k=b+1}^{2b}\varepsilon_k^{(2)} + \cdots + \sum_{k=N-b+1}^N \varepsilon_k^{(N/b)}.
\end{equation}
By this operation, we generate an ensemble of cover times from
processes that have been block shuffled. If the block length $b=N$,
then the original cover process is left intact and no shuffling
occurs. Meanwhile, if $b=1$, then each waiting time $\varepsilon_k$ is
drawn randomly from the ensemble distribution of waiting times to the
$k$th unvisited site. More generally, the block length plays the role
of a ``high-pass" filter that destroys correlations with
characteristic scale larger than $b$. Thus, for $b=1$, the
block-shuffled cover time $C_{N}^{\text{shuff}}(0)$ is a sum of
waiting times from independent realizations. The resulting
$C_{N}^{\text{shuff}}(0)$ could only be non-Gaussian if the
$\varepsilon_k$ were sufficiently nonidentical.

\begin{figure}[h!]
    \centering
    \includegraphics[width=\columnwidth]{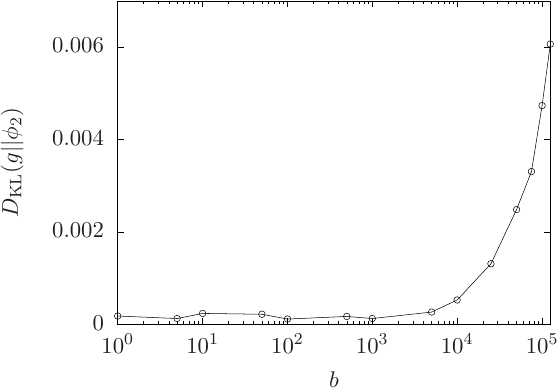}
    \caption{Kullback-Leibler divergence from a Gaussian distribution
      for the rescaled cover time, as a function of block length $b$,
      with $N=50^3$ over an ensemble of $10^5$ realizations.}
    \label{F:shuffle}
\end{figure}

As a measure of discrepancy between the empirical density $\phi_2(z)$
of shuffled cover times and a standard Gaussian density $g(z)$, we
compute the Kullback-Leibler divergence (KL) from $\phi_2(z)$ to $g(z)$,
\begin{equation}
    D_{\text{KL}}(g || \phi_2) = \int \D z \, g(z) \ln\frac{g(z)}{\phi_2(z)}
\end{equation}
for different block lengths $b$. Figure~\ref{F:shuffle} shows that a comparatively
large block length of $b \lesssim 5 \times 10^3 = N/25$ is enough to
recover Gaussian cover time behavior, thus suggesting that long-range
correlations are at play.

\begin{figure}[h!]
    \centering
    \includegraphics[width=\columnwidth]{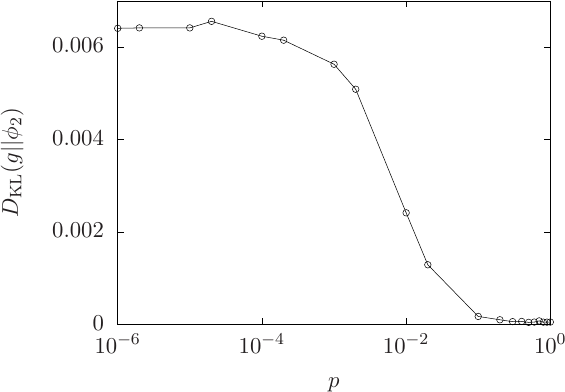}
    \caption{Kullback-Leibler divergence from a Gaussian density for
      the rescaled cover time, as a function of teleportation
      probability $p$, with $N=50^3$ over an ensemble of $10^6$
      independent realizations.}
    \label{F:teleport}
\end{figure}

It is instructive to consider another modification of the cover
process (also implemented in \cite{Chupeau2015} in the unaccelerated
case). Instead of splicing together blocks of cover from independent
realizations, we intermittently allow the random walker to
``teleport'' to a randomly chosen site. Thus, the walker performs a
teleportation jump with probability $p$, and a nearest-neighbor step
with probability $(1-p)$. If $p=0$, the original cover process is
recovered. If $p=1$, the walker effectively explores a fully connected
graph. Figure~\ref{F:teleport} shows that a teleportation probability of
approximately $p \gtrsim 0.1$ is enough to recover Gaussian cover time
behavior.

In conclusion, we attribute the non-Gaussianity of $\alpha = 0$
accelerated cover times in $d=3$ to correlations in the sequence of
sites visited as the lattice is covered. However, we are not able to
explain why such correlations can be ignored for $d\ge 4$, or for
deceleration protocols with $\alpha > 1$.

\section{Conclusion \label{S:conclusion}}

We have studied the cover times of accelerated and decelerated random
walks on a torus in dimensions $d\ge 3$. Building on the work of
Aldous~\cite{Aldous_book1989} and Belius~\cite{Belius2013}, we
conjecture a generalized cover time which agrees well with numerics
for a range of acceleration-deceleration protocols and dimensions. The
$\alpha$-indexed family of cover time distributions are in fact those
describing the roughness of $1/f^{\alpha}$ Gaussian
signals~\cite{Antal2002}, which include Gaussian ($0 \le \alpha \le
1/2$), Gumbel ($\alpha = 1$) and exponential ($\alpha \to \infty$)
distributions, to name a few.

A notable exception to our conjecture is for $\alpha = 0$ in $d=3$,
where we find a cover time distribution somewhat resembling a
Tracy-Widom distribution from the Gaussian orthogonal ensemble of
random matrices. Although the numerics do not support this
identification conclusively, it is interesting to speculate whether a
connection between accelerated cover in $d=3$ and random matrices
exists, e.g. via mappings to Kardar Parisi Zhang
interfaces~\cite{Takeuchi2018}, Gaussian free
fields~\cite{Ding2012,Fyodorov2012}, or spin
glasses~\cite{Castellana2014}.

This study leaves a number of open questions, such as the
identification of the cover time distribution for $\alpha = 0$ in
$d=3$, and why this distribution is particular to $d=3$.

\begin{acknowledgments}
The authors thank Andy Thomas for computer support. G.M. thanks LML and
Imperial College for financial support via the LML-Roth scholarship.
\end{acknowledgments}

\appendix

\section{Roughness of $1/f^{\alpha}$ Gaussian signals}

A $1/f^{\alpha}$ signal $h(x)$ of length $L$ is sampled over $N$
Fourier modes according to
\begin{equation}
    h(x) \propto \sum_{k=1}^N \frac{1}{k^{\alpha/2}} \left[ a_k \sin(2\pi k x/L) + b_k \cos(2\pi k x/L) \right],
\label{E:noise}
\end{equation}
where the amplitudes $a_k$ and $b_k$ are independent standard Gaussian
random variables~\cite{Antal2002}. By construction, the signal is
periodic with zero mean, and its power spectrum decays as
$1/k^{\alpha}$. The integrated power spectrum
\begin{align}
    w^2(\alpha) &\propto \frac{1}{L}\int_0^L \D x \, h^2(x) \\
    &\propto \sum_{k=1}^N \frac{1}{k^{\alpha}}(a_k^2 + b_k^2) 
\end{align}
by Parseval's theorem. Since the sum of two Gaussian squared random
variables is exponentially distributed,
\begin{equation}
    w^2(\alpha) \propto \sum_{k=1}^N \frac{\varepsilon_k}{k^{\alpha}}.   
\end{equation}
Hence, apart from an $N^{\alpha}$ prefactor, the integrated power
spectrum of $1/f^{\alpha}$ signals has the same distribution as the
coupon collection time $C_N(\alpha)$ discussed in the main text.

In the language of interfaces, Eq.~\eqref{E:noise} describes a
periodic steady-state height profile, and the integrated power
spectrum is equivalent to the profile's
roughness~\cite{Barabasi1995}. A review of $1/f^{\alpha}$ signals can
be found in~\cite{Antal2002}.

\section{Lindeberg condition}

Given a collection of independent but not necessarily identical random
variables $\{X_k\}_{k=1}^N$ with (finite) variances
$\{\sigma^2_k\}_{k=1}^N$, the Lindeberg condition~\cite{Resnick1999}
guarantees that their rescaled sum is still Gaussian-distributed,
provided that
\begin{equation}
    \frac{\max\limits_{1 \leq k \leq N}\sigma^2_k}{\sum_{k=1}^N\sigma^2_k}\to 0, \quad N\to \infty.
\label{E:Lindeberg}
\end{equation}

In our context, the collection of random variables
$\varepsilon_k/k^{\alpha}$ have variances $1/k^{2\alpha}$. Therefore,
satisfying Eq.~\eqref{E:Lindeberg} requires
\begin{equation}
    \sum_{k=1}^N \frac{1}{k^{2\alpha}} \to \infty, \quad N\to\infty,
\end{equation}
i.e. that $0 \le \alpha \le 1/2$.


\begin{thebibliography}{48}
\expandafter\ifx\csname natexlab\endcsname\relax\def\natexlab#1{#1}\fi
\expandafter\ifx\csname bibnamefont\endcsname\relax
  \def\bibnamefont#1{#1}\fi
\expandafter\ifx\csname bibfnamefont\endcsname\relax
  \def\bibfnamefont#1{#1}\fi
\expandafter\ifx\csname citenamefont\endcsname\relax
  \def\citenamefont#1{#1}\fi
\expandafter\ifx\csname url\endcsname\relax
  \def\url#1{\texttt{#1}}\fi
\expandafter\ifx\csname urlprefix\endcsname\relax\def\urlprefix{URL }\fi
\providecommand{\bibinfo}[2]{#2}
\providecommand{\eprint}[2][]{\url{#2}}

\bibitem[{\citenamefont{Holst}(1986)}]{Holst1986}
\bibinfo{author}{\bibfnamefont{L.}~\bibnamefont{Holst}},
  \bibinfo{journal}{Int.~Stat.~Rev.} \textbf{\bibinfo{volume}{54}},
  \bibinfo{pages}{15} (\bibinfo{year}{1986}).

\bibitem[{\citenamefont{\'A.~Gonz\'alez and Pacheco}(2005)}]{Gonzalez2005}
\bibinfo{author}{\bibfnamefont{J.~B.~G.} \bibnamefont{\'A.~Gonz\'alez}}
  \bibnamefont{and} \bibinfo{author}{\bibfnamefont{A.~F.}
  \bibnamefont{Pacheco}}, \bibinfo{journal}{Am.~J.~Phys.}
  \textbf{\bibinfo{volume}{73}}, \bibinfo{pages}{946} (\bibinfo{year}{2005}).

\bibitem[{\citenamefont{Turban}(2019)}]{Turban2019}
\bibinfo{author}{\bibfnamefont{L.}~\bibnamefont{Turban}},
  \bibinfo{journal}{J.~Phys.~A-Math.~Gen.} \textbf{\bibinfo{volume}{53}},
  \bibinfo{pages}{035001} (\bibinfo{year}{2019}).

\bibitem[{\citenamefont{B.~Ottino-L\"offler and Strogatz}(2017)}]{Ottino2017}
\bibinfo{author}{\bibfnamefont{J.~G.~S.} \bibnamefont{B.~Ottino-L\"offler}}
  \bibnamefont{and} \bibinfo{author}{\bibfnamefont{S.~H.}
  \bibnamefont{Strogatz}}, \bibinfo{journal}{Phys.~Rev.~E}
  \textbf{\bibinfo{volume}{96}}, \bibinfo{pages}{012313}
  (\bibinfo{year}{2017}).

\bibitem[{\citenamefont{Zoroa et~al.}(2017)\citenamefont{Zoroa, Lesigne,
  Fern{\'a}ndez-S{\'a}ez, Zoroa, and Casas}}]{Zoroa2017}
\bibinfo{author}{\bibfnamefont{N.}~\bibnamefont{Zoroa}},
  \bibinfo{author}{\bibfnamefont{E.}~\bibnamefont{Lesigne}},
  \bibinfo{author}{\bibfnamefont{M.~J.} \bibnamefont{Fern{\'a}ndez-S{\'a}ez}},
  \bibinfo{author}{\bibfnamefont{P.}~\bibnamefont{Zoroa}}, \bibnamefont{and}
  \bibinfo{author}{\bibfnamefont{J.}~\bibnamefont{Casas}},
  \bibinfo{journal}{J.~R.~Soc~Interface} \textbf{\bibinfo{volume}{14}},
  \bibinfo{pages}{20160643} (\bibinfo{year}{2017}).

\bibitem[{\citenamefont{Johnson and Kotz}(1977)}]{JohnsonKotz1977}
\bibinfo{author}{\bibfnamefont{N.~L.} \bibnamefont{Johnson}} \bibnamefont{and}
  \bibinfo{author}{\bibfnamefont{S.}~\bibnamefont{Kotz}},
  \emph{\bibinfo{title}{Urn models and their application}}
  (\bibinfo{publisher}{{John Wiley \& Sons}}, \bibinfo{year}{1977}).

\bibitem[{\citenamefont{Holst}(2001)}]{Holst2001}
\bibinfo{author}{\bibfnamefont{L.}~\bibnamefont{Holst}},
  \bibinfo{journal}{Extremes} \textbf{\bibinfo{volume}{4}},
  \bibinfo{pages}{129} (\bibinfo{year}{2001}).

\bibitem[{\citenamefont{Erd{\H o}s and R{\'e}nyi}(1961)}]{Erdos1961}
\bibinfo{author}{\bibfnamefont{P.}~\bibnamefont{Erd{\H o}s}} \bibnamefont{and}
  \bibinfo{author}{\bibfnamefont{A.}~\bibnamefont{R{\'e}nyi}},
  \bibinfo{journal}{Magyar. Tud Akad. Mat. Kutato Int. K{\"o}zl}
  \textbf{\bibinfo{volume}{6}}, \bibinfo{pages}{215} (\bibinfo{year}{1961}).

\bibitem[{\citenamefont{Aldous}(1983)}]{Aldous1983}
\bibinfo{author}{\bibfnamefont{D.~J.} \bibnamefont{Aldous}},
  \bibinfo{journal}{Z.~Wahrsch.~Verw.~Gebiete} \textbf{\bibinfo{volume}{62}},
  \bibinfo{pages}{361} (\bibinfo{year}{1983}).

\bibitem[{\citenamefont{Aldous}(1989{\natexlab{a}})}]{Aldous1989}
\bibinfo{author}{\bibfnamefont{D.}~\bibnamefont{Aldous}},
  \bibinfo{journal}{J.~Theor.~Probab.} \textbf{\bibinfo{volume}{2}},
  \bibinfo{pages}{87} (\bibinfo{year}{1989}{\natexlab{a}}).

\bibitem[{\citenamefont{Wilf}(1989)}]{Wilf1989}
\bibinfo{author}{\bibfnamefont{H.~S.} \bibnamefont{Wilf}},
  \bibinfo{journal}{Amer.~Math.~Mon.} \textbf{\bibinfo{volume}{96}},
  \bibinfo{pages}{704} (\bibinfo{year}{1989}).

\bibitem[{\citenamefont{Aldous}(1989{\natexlab{b}})}]{Aldous_book1989}
\bibinfo{author}{\bibfnamefont{D.}~\bibnamefont{Aldous}},
  \emph{\bibinfo{title}{{Probability approximations via the Poisson clumping
  heuristic}}} (\bibinfo{publisher}{{Springer-Verlag, New York}},
  \bibinfo{year}{1989}{\natexlab{b}}).

\bibitem[{\citenamefont{Belius}(2013)}]{Belius2013}
\bibinfo{author}{\bibfnamefont{D.}~\bibnamefont{Belius}},
  \bibinfo{journal}{Probab. Theory Relat. Fields}
  \textbf{\bibinfo{volume}{157}}, \bibinfo{pages}{635} (\bibinfo{year}{2013}).

\bibitem[{\citenamefont{Brummelhuis and Hilhorst}(1991)}]{Brummelhuis1991}
\bibinfo{author}{\bibfnamefont{M.~J.~A.~M.} \bibnamefont{Brummelhuis}}
  \bibnamefont{and} \bibinfo{author}{\bibfnamefont{H.~J.}
  \bibnamefont{Hilhorst}}, \bibinfo{journal}{Physica~A}
  \textbf{\bibinfo{volume}{176}}, \bibinfo{pages}{387} (\bibinfo{year}{1991}).

\bibitem[{\citenamefont{Brummelhuis and Hilhorst}(1992)}]{Brummelhuis1992}
\bibinfo{author}{\bibfnamefont{M.~J.~A.~M.} \bibnamefont{Brummelhuis}}
  \bibnamefont{and} \bibinfo{author}{\bibfnamefont{H.~J.}
  \bibnamefont{Hilhorst}}, \bibinfo{journal}{Physica~A}
  \textbf{\bibinfo{volume}{185}}, \bibinfo{pages}{35} (\bibinfo{year}{1992}).

\bibitem[{\citenamefont{Freund and Grassberger}(1993)}]{Freund1993}
\bibinfo{author}{\bibfnamefont{H.}~\bibnamefont{Freund}} \bibnamefont{and}
  \bibinfo{author}{\bibfnamefont{P.}~\bibnamefont{Grassberger}},
  \bibinfo{journal}{Physica~A} \textbf{\bibinfo{volume}{192}},
  \bibinfo{pages}{465} (\bibinfo{year}{1993}).

\bibitem[{\citenamefont{Sznitman}(2012)}]{Sznitman2012}
\bibinfo{author}{\bibfnamefont{A.-S.} \bibnamefont{Sznitman}},
  \bibinfo{journal}{Ann.~Probab.} \textbf{\bibinfo{volume}{40}},
  \bibinfo{pages}{2400} (\bibinfo{year}{2012}).

\bibitem[{\citenamefont{Drewitz et~al.}(2014)\citenamefont{Drewitz, R\'{a}th,
  and Sapozhnikov}}]{Drewitz2014}
\bibinfo{author}{\bibfnamefont{A.}~\bibnamefont{Drewitz}},
  \bibinfo{author}{\bibfnamefont{B.}~\bibnamefont{R\'{a}th}}, \bibnamefont{and}
  \bibinfo{author}{\bibfnamefont{A.}~\bibnamefont{Sapozhnikov}},
  \emph{\bibinfo{title}{{An Introduction to Random Interlacements}}}
  (\bibinfo{publisher}{{Springer Briefs in Mathematics, Springer}},
  \bibinfo{year}{2014}).

\bibitem[{\citenamefont{Antal et~al.}(2002)\citenamefont{Antal, Droz,
  Gy\"orgyi, and R\'acz}}]{Antal2002}
\bibinfo{author}{\bibfnamefont{T.}~\bibnamefont{Antal}},
  \bibinfo{author}{\bibfnamefont{M.}~\bibnamefont{Droz}},
  \bibinfo{author}{\bibfnamefont{G.}~\bibnamefont{Gy\"orgyi}},
  \bibnamefont{and} \bibinfo{author}{\bibfnamefont{Z.}~\bibnamefont{R\'acz}},
  \bibinfo{journal}{Phys.~Rev.~E} \textbf{\bibinfo{volume}{65}},
  \bibinfo{pages}{046140} (\bibinfo{year}{2002}).

\bibitem[{\citenamefont{Coquel et~al.}(2013)\citenamefont{Coquel, Jacob,
  Primet, Demarez, Dimiccoli, Julou, Moisan, Lindner, and Berry}}]{Coquel2013}
\bibinfo{author}{\bibfnamefont{A.-S.} \bibnamefont{Coquel}},
  \bibinfo{author}{\bibfnamefont{J.-P.} \bibnamefont{Jacob}},
  \bibinfo{author}{\bibfnamefont{M.}~\bibnamefont{Primet}},
  \bibinfo{author}{\bibfnamefont{A.}~\bibnamefont{Demarez}},
  \bibinfo{author}{\bibfnamefont{M.}~\bibnamefont{Dimiccoli}},
  \bibinfo{author}{\bibfnamefont{T.}~\bibnamefont{Julou}},
  \bibinfo{author}{\bibfnamefont{L.}~\bibnamefont{Moisan}},
  \bibinfo{author}{\bibfnamefont{A.~B.} \bibnamefont{Lindner}},
  \bibnamefont{and} \bibinfo{author}{\bibfnamefont{H.}~\bibnamefont{Berry}},
  \bibinfo{journal}{PLOS~Comput.~Biol.} \textbf{\bibinfo{volume}{9}},
  \bibinfo{pages}{1} (\bibinfo{year}{2013}).

\bibitem[{\citenamefont{Hidalgo-Soria and Barkai}(2019)}]{Soria2019}
\bibinfo{author}{\bibfnamefont{M.}~\bibnamefont{Hidalgo-Soria}}
  \bibnamefont{and} \bibinfo{author}{\bibfnamefont{E.}~\bibnamefont{Barkai}}
  (\bibinfo{year}{2019}), \bibinfo{note}{{arXiv:1909.07189v2}}.

\bibitem[{\citenamefont{Benichou et~al.}(2011)\citenamefont{Benichou, Loverdo,
  Moreau, and Voituriez}}]{Benichou2011}
\bibinfo{author}{\bibfnamefont{O.}~\bibnamefont{Benichou}},
  \bibinfo{author}{\bibfnamefont{C.}~\bibnamefont{Loverdo}},
  \bibinfo{author}{\bibfnamefont{M.}~\bibnamefont{Moreau}}, \bibnamefont{and}
  \bibinfo{author}{\bibfnamefont{R.}~\bibnamefont{Voituriez}},
  \bibinfo{journal}{Rev.~Mod.~Phys} \textbf{\bibinfo{volume}{83}},
  \bibinfo{pages}{81} (\bibinfo{year}{2011}).

\bibitem[{\citenamefont{Leadbetter et~al.}(1983)\citenamefont{Leadbetter,
  Lindgren, and Rootz\'{e}n}}]{Leadbetter1983}
\bibinfo{author}{\bibfnamefont{M.~R.} \bibnamefont{Leadbetter}},
  \bibinfo{author}{\bibfnamefont{G.}~\bibnamefont{Lindgren}}, \bibnamefont{and}
  \bibinfo{author}{\bibfnamefont{H.}~\bibnamefont{Rootz\'{e}n}},
  \emph{\bibinfo{title}{{Extremes and Related Properties of Random Sequences
  and Processes}}} (\bibinfo{publisher}{{Springer-Verlag, New York, Heidelberg,
  Berlin}}, \bibinfo{year}{1983}).

\bibitem[{\citenamefont{Cox and Isham}(1980)}]{Cox1980}
\bibinfo{author}{\bibfnamefont{D.}~\bibnamefont{Cox}} \bibnamefont{and}
  \bibinfo{author}{\bibfnamefont{V.}~\bibnamefont{Isham}},
  \emph{\bibinfo{title}{{Point processes}}} (\bibinfo{publisher}{{CRC Press}},
  \bibinfo{year}{1980}).

\bibitem[{\citenamefont{Weissman}(1988)}]{Weissman1988}
\bibinfo{author}{\bibfnamefont{M.~B.} \bibnamefont{Weissman}},
  \bibinfo{journal}{Rev.~Mod.~Phys.} \textbf{\bibinfo{volume}{60}},
  \bibinfo{pages}{537} (\bibinfo{year}{1988}).

\bibitem[{\citenamefont{Edwards and Wilkinson}(1982)}]{Edwards1982}
\bibinfo{author}{\bibfnamefont{S.~F.} \bibnamefont{Edwards}} \bibnamefont{and}
  \bibinfo{author}{\bibfnamefont{D.~R.} \bibnamefont{Wilkinson}},
  \bibinfo{journal}{Proc.~R.~Soc.~Lond. A}
  \textbf{\bibinfo{volume}{381}}, \bibinfo{pages}{17} (\bibinfo{year}{1982}).

\bibitem[{\citenamefont{Mullins}(1957)}]{Mullins1957}
\bibinfo{author}{\bibfnamefont{W.~W.} \bibnamefont{Mullins}},
  \bibinfo{journal}{J.~Appl.~Phys.} \textbf{\bibinfo{volume}{28}},
  \bibinfo{pages}{333} (\bibinfo{year}{1957}).

\bibitem[{\citenamefont{Kolmogorov}(1933)}]{Kolmogorov1933}
\bibinfo{author}{\bibfnamefont{A.~N.} \bibnamefont{Kolmogorov}},
  \bibinfo{journal}{Giorn.~Ist.~Ital.~Attuari} \textbf{\bibinfo{volume}{4}},
  \bibinfo{pages}{1} (\bibinfo{year}{1933}).

\bibitem[{\citenamefont{Foltin et~al.}(1994)\citenamefont{Foltin, Oerding,
  R\'{a}cz, Workman, and Zia}}]{Foltin1994}
\bibinfo{author}{\bibfnamefont{G.}~\bibnamefont{Foltin}},
  \bibinfo{author}{\bibfnamefont{K.}~\bibnamefont{Oerding}},
  \bibinfo{author}{\bibfnamefont{Z.}~\bibnamefont{R\'{a}cz}},
  \bibinfo{author}{\bibfnamefont{R.~L.} \bibnamefont{Workman}},
  \bibnamefont{and} \bibinfo{author}{\bibfnamefont{R.~K.~P.}
  \bibnamefont{Zia}}, \bibinfo{journal}{Phys.~Rev.~E}
  \textbf{\bibinfo{volume}{50}}, \bibinfo{pages}{R639} (\bibinfo{year}{1994}).

\bibitem[{\citenamefont{Biane et~al.}(2001)\citenamefont{Biane, Pitman, and
  Yor}}]{Biane2001}
\bibinfo{author}{\bibfnamefont{P.}~\bibnamefont{Biane}},
  \bibinfo{author}{\bibfnamefont{J.}~\bibnamefont{Pitman}}, \bibnamefont{and}
  \bibinfo{author}{\bibfnamefont{M.}~\bibnamefont{Yor}},
  \bibinfo{journal}{Bull.~Amer.~Math.~Soc.~(N.S.)}
  \textbf{\bibinfo{volume}{38}}, \bibinfo{pages}{435} (\bibinfo{year}{2001}).

\bibitem[{\citenamefont{Font-Clos and Moloney}(2016)}]{Font2016}
\bibinfo{author}{\bibfnamefont{F.}~\bibnamefont{Font-Clos}} \bibnamefont{and}
  \bibinfo{author}{\bibfnamefont{N.~R.} \bibnamefont{Moloney}},
  \bibinfo{journal}{Phys.~Rev.~E} \textbf{\bibinfo{volume}{94}},
  \bibinfo{pages}{{030102(R)}} (\bibinfo{year}{2016}).

\bibitem[{\citenamefont{Botet and P{\l}oszajczak}(2005)}]{Botet2005}
\bibinfo{author}{\bibfnamefont{R.}~\bibnamefont{Botet}} \bibnamefont{and}
  \bibinfo{author}{\bibfnamefont{M.}~\bibnamefont{P{\l}oszajczak}},
  \bibinfo{journal}{Phys.~Rev.~Lett.} \textbf{\bibinfo{volume}{95}},
  \bibinfo{pages}{185702} (\bibinfo{year}{2005}).

\bibitem[{\citenamefont{Watson}(1961)}]{Watson1961}
\bibinfo{author}{\bibfnamefont{G.~S.} \bibnamefont{Watson}},
  \bibinfo{journal}{Biometrika} \textbf{\bibinfo{volume}{48}},
  \bibinfo{pages}{109} (\bibinfo{year}{1961}).

\bibitem[{\citenamefont{M.~Plischke and Zia}(1994)}]{Plischke1994}
\bibinfo{author}{\bibfnamefont{Z.~R.} \bibnamefont{M.~Plischke}}
  \bibnamefont{and} \bibinfo{author}{\bibfnamefont{R.~K.~P.}
  \bibnamefont{Zia}}, \bibinfo{journal}{Phys.~Rev.~E}
  \textbf{\bibinfo{volume}{50}}, \bibinfo{pages}{3589} (\bibinfo{year}{1994}).

\bibitem[{\citenamefont{Aldous and Fill}()}]{AldousFill}
\bibinfo{author}{\bibfnamefont{D.}~\bibnamefont{Aldous}} \bibnamefont{and}
  \bibinfo{author}{\bibfnamefont{J.~A.} \bibnamefont{Fill}},
  \emph{\bibinfo{title}{{Reversible Markov Chains and Random Walks on
  Graphs}}},
  \bibinfo{howpublished}{\url{http://www.stat.berkeley.edu/users/aldous/RWG/book.html}},
  \bibinfo{note}{in preparation}.

\bibitem[{\citenamefont{Zlatanov and Kocarev}(2009)}]{Zlatanov2009}
\bibinfo{author}{\bibfnamefont{N.}~\bibnamefont{Zlatanov}} \bibnamefont{and}
  \bibinfo{author}{\bibfnamefont{L.}~\bibnamefont{Kocarev}},
  \bibinfo{journal}{Phys.~Rev.~E} \textbf{\bibinfo{volume}{80}},
  \bibinfo{pages}{041102} (\bibinfo{year}{2009}).

\bibitem[{\citenamefont{Maier and Brockmann}(2017)}]{Maier2017}
\bibinfo{author}{\bibfnamefont{B.~F.} \bibnamefont{Maier}} \bibnamefont{and}
  \bibinfo{author}{\bibfnamefont{D.}~\bibnamefont{Brockmann}},
  \bibinfo{journal}{Phys.~Rev.~E} \textbf{\bibinfo{volume}{96}},
  \bibinfo{pages}{042307} (\bibinfo{year}{2017}).

\bibitem[{\citenamefont{Imhof}(1985)}]{Imhof1985}
\bibinfo{author}{\bibfnamefont{J.~P.} \bibnamefont{Imhof}},
  \bibinfo{journal}{Ann.~Probab.} \textbf{\bibinfo{volume}{13}},
  \bibinfo{pages}{1011} (\bibinfo{year}{1985}).

\bibitem[{\citenamefont{Grassberger}(2017)}]{Grassberger2017}
\bibinfo{author}{\bibfnamefont{P.}~\bibnamefont{Grassberger}},
  \bibinfo{journal}{Phys.~Rev.~E} \textbf{\bibinfo{volume}{96}},
  \bibinfo{pages}{012115} (\bibinfo{year}{2017}).

\bibitem[{\citenamefont{Belius and Kistler}(2017)}]{Belius2017}
\bibinfo{author}{\bibfnamefont{D.}~\bibnamefont{Belius}} \bibnamefont{and}
  \bibinfo{author}{\bibfnamefont{N.}~\bibnamefont{Kistler}},
  \bibinfo{journal}{Probab. Theory Relat. Fields}
  \textbf{\bibinfo{volume}{167}}, \bibinfo{pages}{461} (\bibinfo{year}{2017}).

\bibitem[{\citenamefont{Chupeau et~al.}(2015)\citenamefont{Chupeau,
  B{\'e}nichou, and Voituriez}}]{Chupeau2015}
\bibinfo{author}{\bibfnamefont{M.}~\bibnamefont{Chupeau}},
  \bibinfo{author}{\bibfnamefont{O.}~\bibnamefont{B{\'e}nichou}},
  \bibnamefont{and}
  \bibinfo{author}{\bibfnamefont{R.}~\bibnamefont{Voituriez}},
  \bibinfo{journal}{Nat.~Phys.} \textbf{\bibinfo{volume}{11}},
  \bibinfo{pages}{844} (\bibinfo{year}{2015}).

\bibitem[{\citenamefont{Glasser and Zucker}(1977)}]{Glasser1977}
\bibinfo{author}{\bibfnamefont{M.~L.} \bibnamefont{Glasser}} \bibnamefont{and}
  \bibinfo{author}{\bibfnamefont{I.~J.} \bibnamefont{Zucker}},
  \bibinfo{journal}{Proc.~Natl.~Acad.~Sci.~USA} \textbf{\bibinfo{volume}{74}},
  \bibinfo{pages}{1800} (\bibinfo{year}{1977}).

\bibitem[{\citenamefont{Takeuchi}(2018)}]{Takeuchi2018}
\bibinfo{author}{\bibfnamefont{K.~A.} \bibnamefont{Takeuchi}},
  \bibinfo{journal}{Physica A} \textbf{\bibinfo{volume}{504}},
  \bibinfo{pages}{77} (\bibinfo{year}{2018}).

\bibitem[{\citenamefont{Ding et~al.}(2012)\citenamefont{Ding, Lee, and
  Peres}}]{Ding2012}
\bibinfo{author}{\bibfnamefont{J.}~\bibnamefont{Ding}},
  \bibinfo{author}{\bibfnamefont{J.~R.} \bibnamefont{Lee}}, \bibnamefont{and}
  \bibinfo{author}{\bibfnamefont{Y.}~\bibnamefont{Peres}},
  \bibinfo{journal}{Ann.~Math.} \textbf{\bibinfo{volume}{175}},
  \bibinfo{pages}{1409} (\bibinfo{year}{2012}).

\bibitem[{\citenamefont{Fyodorov and Nadal}(2012)}]{Fyodorov2012}
\bibinfo{author}{\bibfnamefont{Y.~V.} \bibnamefont{Fyodorov}} \bibnamefont{and}
  \bibinfo{author}{\bibfnamefont{C.}~\bibnamefont{Nadal}},
  \bibinfo{journal}{Phys.~Rev.~Lett.} \textbf{\bibinfo{volume}{109}},
  \bibinfo{pages}{167203} (\bibinfo{year}{2012}).

\bibitem[{\citenamefont{Castellana}(2014)}]{Castellana2014}
\bibinfo{author}{\bibfnamefont{M.}~\bibnamefont{Castellana}},
  \bibinfo{journal}{Phys.~Rev.~Lett.} \textbf{\bibinfo{volume}{112}},
  \bibinfo{pages}{215701} (\bibinfo{year}{2014}).

\bibitem[{\citenamefont{Barabasi and Stanley}(1995)}]{Barabasi1995}
\bibinfo{author}{\bibfnamefont{A.~L.} \bibnamefont{Barabasi}} \bibnamefont{and}
  \bibinfo{author}{\bibfnamefont{H.~E.} \bibnamefont{Stanley}},
  \emph{\bibinfo{title}{Fractal concepts in surface growth}}
  (\bibinfo{publisher}{Cambridge University Press}, \bibinfo{year}{1995}).

\bibitem[{\citenamefont{Resnick}(1999)}]{Resnick1999}
\bibinfo{author}{\bibfnamefont{S.~I.} \bibnamefont{Resnick}},
  \emph{\bibinfo{title}{A probability path}} (\bibinfo{publisher}{Birkh\"auser,
  Boston}, \bibinfo{year}{1999}).

\end{thebibliography}
\end{document}